\documentstyle[preprint,aps]{revtex}

\begin{document}

\preprint{DUKE-TH-95-94}

\draft

\title{Reheating after Supercooling \\
in the Chiral Phase Transition}

\author{Stanis\l aw Mr\' owczy\' nski}

\address{High-Energy Department, So\l tan Institute for Nuclear Studies,\\
ul. Ho\.za 69, PL - 00-681 Warsaw, Poland}

\author{Berndt M\"uller}

\address{Department of Physics, Duke University, Durham, N.C. 27708-0305, USA}

\date{Revised: 30 August 1995}

\maketitle

\begin{abstract}
We study the conversion of the latent heat of a supercooled quark-gluon
plasma into excitations of the pion field. Supercooling is predicted to
occur in the late stages of the evolution of a quark-gluon plasma
produced in energetic heavy-ion collisions, if the QCD phase transition
is of first order.  The supercooled, chirally symmetric state, which
contains potential energy associated with an energetically unfavorable
field configuration, eventually rolls down to the true minimum of the
effective chiral potential.
When this motion is described in terms of the linear sigma-model,
we find that the energy of the coherent $\sigma-$field is very
efficiently converted into pionic excitations due to anharmonic
oscillations around the minimum. The system is expected to
partially thermalize before its disintegration.
\end{abstract}


The process of hadronization being of essentially nonperturbative
character is not well understood at present. This leads to large
theoretical uncertainties in describing the late stages of temporal
evolution of the quark-gluon plasma predicted to be produced in very
energetic heavy ion collisions.  Recently, the hadronization of an
extended quark-gluon plasma has been studied within the nucleation
model \cite{Cse,CK,Sve} by considering the growth of hadronic bubbles
in the quark-gluon plasma undergoing a first-order phase transition.
Thermal equilibrium between both phases was assumed.  The authors
of refs. \cite{Cse,CK} found that  the conversion of the quark-gluon
plasma into hadrons is {\it slow} with a characteristic time much
greater than 10 fm/c, resulting in significant supercooling of the
plasma phase.  According to these model calculations the system
may cool to $T\approx 0.7T_c$ before hadronizing, corresponding to a
reduction by a factor three in the entropy density.  Of course, the
total entropy remains (at least) constant; the falling entropy density
is compensated by spatial expansion while the system cools below $T_c$.

When the transition to the hadronic phase finally occurs it is, in
general, accompanied by significant reheating due to the release of
the latent heat associated with the first-order phase transition.
Additional entropy may be created in this process, but not necessarily
so \cite{CK}.  If the chirally symmetric phase is supercooled to just
such an extent that its entropy content at the temperature $T_Q<T_c$
exactly equals the entropy content of the equilibrated hadronic phase
at the higher temperature $T_H$ reached after reheating, no entropy
is generated.  If the symmetric phase is supercooled further, then
additional entropy is produced.

The present paper is an attempt to analyze the dynamics of this
reheating process within the context of an effective Lagrangian for
the chiral phase transition.  It is inspired by the recent work by Kofman,
Linde, and Starobinsky \cite{Kof}, who studied the reheating after
inflation in the early Universe.  We here discuss a nonequilibrium
scenario, where the hadronization coincides with the chiral phase
transition.  Our picture is closely related to that of Rajagopal and
Wilczek \cite{Raj} who studied the excitation of long-wavelength modes during
the decay of a supercooled, false chiral vacuum state.  However, we are
here concerned not with the formation of coherent field domains, which
has been numerically studied extensively \cite{Raj,Gav,Asakawa}, but
rather with the excitation of quasi-thermal modes of the pion field.

When the quark-gluon plasma is supercooled, the system is
in a moderately excited state over a chirally symmetric false vacuum.
It does not matter for our purposes whether the entropy carrying modes
in this state are best described by nearly massless quarks or by
hadrons which are light due to chiral symmetry restoration, as some
have suggested \cite{Rho+Brown}.  What is important is that the number
of light modes is much larger in the supercooled, chirally symmetric
state than in the hadronic state with broken chiral symmetry into
which it decays.

Once the instability is triggered, the system rolls down to the true
chirally asymmetric minimum of the effective potential. Since the
oscillations around this minimum are anharmonic, the coherent field is
converted into pionic excitations. As we argue below, the conversion
mechanism is fast and strongly couples the long wavelength modes to
those of short wavelength.  The pionic excitations created by this
mechanism are expected to (partially) thermalize before freeze-out.
However, we emphasize that these pionic excitations do not necessarily
increase the total entropy of the system significantly.  In fact, our
model makes no statement at all about an entropy increase.  Because we
do not explicitly consider the process of decoherence of the excited
pionic modes they carry, strictly speaking, no entropy.  It appears
quite likely that during the decoherence process entropy is being
transferred from other hadronic modes into the originally coherent
excitations of the pion field.  What our model does illuminate, however,
is the mechanism by which the latent heat of the vacuum is converted into
excitations of the pion field and on which time scale this happens.

In the following, we analyze this hadronization mechanism within
the linear sigma model defined by the effective Lagrangian
\begin{equation}\label{sigma}
{\cal L}(t,x) =
{1 \over 2} \partial ^{\mu}{\phi}_a\partial _{\mu} {\phi}_a
-{\lambda \over 4} ({\phi}_a {\phi}_a - v^2)^2 + m_{\pi}^2 v\chi
\end{equation}
where ${\phi}_a \equiv (\chi, \vec\pi)$
and the parameters are $\lambda \approx 20$, $v \approx 90$ MeV,
$m_{\pi} \approx 140$ MeV.  It is important to stress that in the deconfined,
chirally symmetric phase the fields ${\phi}_a$  are treated as an
effective description of quarks and gluons.  The ${\phi}_a-$field is
thought to dominate the low-energy dynamics of quarks and gluons
near the transition, see e.g. \cite{Raj}.  Below the critical temperature
the chiral symmetry is broken and one deals with the {\it physical} fields
$\sigma = v_0 - \chi$ and $\vec\pi$,  which correspond to the sigma
mesons and pions, respectively.  Here $v_0 \approx v$ is the minimum
of the potential in (\ref{sigma}).

When the system is significantly supercooled in the symmetric chiral
phase, where $\sigma \approx v$ and $\vec\pi \approx 0$, the energy
density is dominated by the potential energy
$$
\varepsilon_{\rm pot} = {\lambda \over 4} v^4 \;.
$$
If this energy is converted into thermal energy of the pion gas, i.e.
$$
\varepsilon_{\rm pot} = \varepsilon_{\rm th} = {\pi^2 \over 10} \, T^4 \;,
$$
where the pion mass is neglected,\footnote{Since the average thermal
energy of a massless pion is $3T$, the zero mass or ultrarelativistic
approximation is not bad even for temperatures of the order of the pion
mass.} one finds the temperature $T \approx 135$ MeV, which is close to
the freeze-out temperature of pions observed in relativistic
nucleus-nucleus collisions \cite{Ags}.  Thus, even the complete
supercooling of the system with subsequent reheating is not in conflict
with the data. Let us then discuss the scenario in more detail.

The sigma and pion fields satisfy the following equations of motion:
\begin{equation}
\Big[ \partial^2 + 2 \lambda v^2 - 3 \lambda v \sigma
+ \lambda \sigma^2 + \lambda \vec\pi^2 \Big] \sigma
=  \lambda v \vec\pi^2  \;, \label{1a}
\end{equation}
\begin{equation}
\Big[ \partial^2 + m^2_{\pi} - 2  \lambda v \sigma + \lambda \sigma^2
+ \lambda \vec\pi^2  \Big] \vec\pi = 0 \;. \label{1b}
\end{equation}
We use these equations to discuss the temporal evolution of the system
which initially is in the chirally symmetric phase. The fields are treated
as classical. Since the system is assumed to be significantly supercooled,
the initial values of the fields are taken as $\sigma \approx v$ and
$\vec\pi \approx 0$. Then, the system appears on the top of a ``Mexican
hat'' potential.  When the $\sigma-$field is rolling down to the potential
minimum at $\sigma=0$, its amplitude decreases. Therefore one expects
that $v^2 \sigma > v \sigma^2 > \sigma^3 $, and as a zeroth approximation,
we neglect in eqs.~(\ref{1a}, \ref{1b}) the terms which are quadratic and
cubic in the fields. In this way we get the equations
\begin{equation}
\Big[ \partial^2  + m_{\sigma}^2 \Big] \sigma^{(0)} = 0 \;,
\qquad \left[\partial^2+m^2_{\pi}\right] \vec\pi^{(0)} = 0 \;,
\label{2}
\end{equation}
with $m_{\sigma} \equiv  \sqrt{2 \lambda }v = 600$ MeV, which have plane
wave solutions.  Keeping in mind the initial conditions, we choose
the solutions of eqs.~(\ref{2}) as
\begin{equation}\label{zeroth}
\sigma^{(0)} = \sigma_0 {\rm cos}(m_{\sigma}t + \varphi)\;,
\qquad \vec\pi^{(0)} = 0 \;, \label{3}
\end{equation}
where the field $\sigma^{(0)}$ is assumed to be homogeneous.  The
solution (\ref{3}) describes undamped collective oscillations of
the $\sigma -$field around its minimum.  Now we substitute the solutions
(\ref{3}) into eqs. (\ref{1a}, \ref{1b}) and keep the terms no more
than quadratic in the $\sigma -$field and linear in the $\vec\pi -$field.
Changing the time variable $(2z \equiv m_{\sigma}t + \varphi)$ and
writing down the equations for modes labeled by the momentum ${\bf k}$
we get
\begin{equation}
\left[ {d^2 \over dz^2} + A_{\sigma}
- 2 q_{\sigma}{\rm cos}(2z) \right] \sigma^{(1)}_{\bf k}(z) = 0 \;,
\label{4a}
\end{equation}
\begin{equation}
\left[ {d^2 \over dz^2} + A_{\pi} - 2 q_{\pi}{\rm cos}(2z) \right]
\vec\pi^{(1)}_{\bf k}(z) = 0 \;,
\label{4b}
\end{equation}
where
\begin{equation}
A_{\sigma} \equiv 4{m_{\sigma}^2 + {\bf k}^2 \over m_{\sigma}^2}
\;, \qquad
q_{\sigma} \equiv 3 {\sigma_0 \over v} \;,
\label{5a}
\end{equation}
and
\begin{equation}
A_{\pi} \equiv 4{ m^2_{\pi}+{\bf k}^2 \over m_{\sigma}^2}
\;,\qquad q_{\pi} \equiv 2 {\sigma_0 \over v} \;.
\label{5b}
\end{equation}
Eqs.~(\ref{4a},\ref{4b}) are written down in the form of the well
known Mathieu equation \cite{Abr}, which corresponds to the wave equation
in elliptic coordinates.  As discussed above $\sigma_0 < v$, thus
$0< q_{\sigma}< 3$ and $0< q_{\pi}< 2$, while $A_{\sigma}\ge 4$ and
$A_{\pi}\ge 0.2$ .

The solutions of Mathieu's equation are of the form
$$
F_{\nu}(z) = e^{i\nu z} P(z) \;,
$$
where the constant $\nu$, which is called the characteristic
exponent, depends on $A$ and $q$ and $P(z)$ is a periodic function
with the period $\pi$.  It is easy to see that for an infinitesimally
small $q$ there are narrow regions of resonant solutions of the equations
(\ref{4a},\ref{4b}) for $A \approx N^2$ where $N$ is an arbitrary integer
number. Then, the characteristic frequency of the wave equation is a
multiple of the frequency of the stimulating force.  Let us consider
for illustration the first resonant region around $A = 1$. The
parameters $A$, $q$ and $\nu$ are then connected by the equation \cite{Abr}
$$
A = \nu^2 + {q^2 \over 2(\nu^2 -1)} \;, \quad {\rm for} \;
\nu \not= 1 \;.
$$
For $A=1$ one immediately finds
$$
\nu^2 = 1 \pm {i \over \sqrt{2}} \, q \;.
$$
Decomposing $\nu$ into the real and imaginary parts
with $\vert{\rm Re}\; \nu \vert \gg \vert{\rm Im}\;\nu \vert $, we get
$$
{\rm Im}\;\nu \approx  \pm {i \over 2\sqrt{2}} \, q \;.
$$
As expected, we have found an unstable mode which generally grows
exponentially in time.

When the amplitude of the stimulating force represented by $q$ increases,
the narrow resonances change into wide resonant bands in the $A-q$
plane. The imaginary part of the characteristic exponent $\nu$
is negative there and the solutions of Mathieu's equation are unstable.
For example, when $q = 1$ the resonant regions extend for $A$ from the
intervals (0, 1.9), (3.9, 4.4), (9.06, 9.08) etc.  As seen, the first
two resonances are rather broad while the third and higher ones remain
narrow for $q = 1$.

Although our zeroth order solution (\ref{zeroth}) is homogeneous,
i.e. ${\bf k} = 0$, it effectively couples (via eqs.~(\ref{4a}, \ref{4b}))
not only to the long wavelength pion modes but also to those
with $\vert {\bf k} \vert \approx m_{\sigma}$. Let us consider how fast
these modes grow.  For a given value of $A$ and $q$, the characteristic
exponents can be read from the charts presented in \cite{Abr}. For
$q_{\sigma} = 3$ and $q_{\pi} = 2$, which are the maximal values, one
finds for $A_{\sigma} \approx A_{\pi} \approx 4$ the imaginary
characteristic exponents equal 0.35 and 0.17, respectively. Then, the
particle numbers proportional to $\vert {\sigma}^{(1)}_{\bf k}\vert^2$
and $\vert\vec\pi^{(1)}_{\bf k} \vert^2$ grow in time as $e^{t/\tau}$
with $\tau_{\sigma} = 0.9 $ fm/c and $\tau_{\pi} = 1.9$ fm/c, where
$\tau^{-1} = m_{\sigma}{\rm Im}\;\nu$.  The characteristic time
$\tau_{\pi}$ of growth of the occupation of single-particle pion modes
appears much smaller than the nucleation time found in \cite{Cse}.
Therefore, the mechanism described here can indeed be responsible for
hadronization of the quark-gluon plasma.

The hadronization which proceeds simultaneously with the transition
from the chirally symmetric to asymmetric state has been discussed in
\cite{Mis}. The hadronization time has been identified there with the
characteristic time of rolling down from the top of the ``Mexican hat''
to the potential minimum, which is of order 0.5 fm/c. In our opinion,
we deal with ``physical'' pions only when the coherence of the $\sigma$
and $\vec\pi$ fields breaks down due to the anharmonic oscillations
around the true vacuum state. Therefore, we identify the hadronization
time with $\tau_{\pi}$ found above which equals about 2 fm/c.

An important feature of our hadronization scenario is the strong coupling
of the soft coherent modes to those with $\vert {\bf k} \vert \approx
m_{\sigma}$.  This implies that the spectrum of produced pions is
rather broad, similar to a thermal spectrum at $T\approx m_{\pi}$.
Therefore, we expect at least partial thermalization of the system.
The reheating or even overheating due to the rapid release of the latent
heat has also been advocated \cite{Cso} within the nucleation model
\cite{Cse}. The thermal spectrum of hadrons observed in relativistic
heavy-ion collisions supports such an expectation.  It might be of
critical importance for the so-called Disoriented Chiral Condensates
actively discussed recently \cite{Raj,Gav,Ans}.

It has been suggested that coherent domains of the pion field can appear
in the nonequilibrium chiral phase transition of the quark-gluon system
created in high energy collisions.  The phenomenon, analogous to the
formation of misaligned domains in a ferromagnet, would be observed
by the coherent pion emission when the domains relax to the ground state
with vanishing pion field.  Specifically, one expects significant
fluctuations in the number of neutral and charged pions. The soft pions
are expected to be particularly sensitive to the domain formation.
Numerical simulations of the linear sigma model represented by the
Lagrangian (\ref{sigma}) have shown \cite{Raj,Gav} that the disoriented
condensates can indeed appear under favorable conditions.

As the reheating proceeds in the hadronizing system, it will generate
additional background on which it is more difficult to observe the
pions from coherent domains.  We note that some of the numerical
simulations \cite{Raj,Gav} do not fully take into account the short
wavelength modes excited by eqs. (\ref{4a},\ref{4b}) due to the finite
lattice spacing, typically of 1 fm.  In the work of Asakawa et al.
\cite{Asakawa} where the lattice spacing is 0.25 fm, there are
indications that reheating occurs.

In summary, if the quark-gluon plasma produced in relativistic heavy-ion
collisions is significantly supercooled  at the late stage of its
evolution, its thermal energy is converted into potential energy of the
$\sigma-$field. The system then rolls down from the chirally symmetric
to the asymmetric state and the hadronization coincides with the chiral
phase transition.  The physical pions emerge when the system
anharmonically oscillates around the true vacuum.  Due to the
efficient resonant coupling between the soft and hard modes,
the system is expected to thermalize again, at least partially, on a
short time scale.

\acknowledgements

We thank the European Centre for Theoretical Studies in Nuclear Physics
and Related Areas in Trento (Italy) for warm hospitality.  This work was
supported, in part, by the U.S. Department of Energy and the Polish
Committee of Scientific Research under grants DE-FG05-90ER40592 and
2-P03B-195-09, respectively.  The authors thank X. N. Wang for an
interesting discussion.

\end{document}